\documentclass[aps,prb,twocolumn,showpacs,ams]{revtex4}

\usepackage{graphicx}
\usepackage{bm}
\usepackage{amssymb} 

\begin{document}

\title{F\"{o}rster energy transfer signatures in optically driven quantum dot molecules}

\author{Juan E. Rolon}
\email{rolon@phy.ohiou.edu}
\affiliation{Department of Physics
and Astronomy and Nanoscale and Quantum Phenomena Institute, Ohio
University, Athens, Ohio 45701-2979}

\author{Sergio E. Ulloa}
\affiliation{Department of Physics and Astronomy and Nanoscale and
Quantum Phenomena Institute, Ohio University, Athens, Ohio
45701-2979}

\date{\today}

\begin{abstract}
The F\"{o}rster resonant energy transfer mechanism (FRET) is investigated in optically driven and electrically gated tunnel coupled quantum dot molecules. Two novel FRET induced optical signatures are found in the dressed excitonic spectrum.  This is constructed from exciton level occupation as function of pump laser energy and applied bias, resembling a level anticrossing spectroscopy measurement. We observe a redistribution of spectral weight and splitting of the exciton  spectral lines. FRET among single excitons induces a splitting in the spatially-direct exciton lines, away from the  anticrossing due to charge tunneling in the molecule. However, near the anticrossing, a novel signature appears as a weak satellite line following an indirect exciton line. FRET signatures may also occur among indirect excitons, appearing as split indirect lines.  In that case, the signatures appear also in the direct biexciton states, as the indirect satellite mixes in near the tunneling anticrossing region.
\end{abstract}

\maketitle

\newcommand{\be}   {\begin{equation}}
\newcommand{\ee}   {\end{equation}}
\newcommand{\ba}   {\begin{eqnarray}}
\newcommand{\ea}   {\end{eqnarray}}
\newcommand{\maxim}   {\mbox{\scriptsize max}}
\newcommand{\reduced}   {\mbox{\scriptsize red}}
\newcommand{\pol}   {\mbox{\scriptsize pol}}
\newcommand{\imp}{\mbox{\scriptsize imp}}

\section{Introduction}
\label{sec: Intro}

Semiconductor quantum dot molecules (QDMs) exhibit remarkable electronic and optical properties determined by their size, shape and material composition.\cite{Bayer,Krenner,Stinaff,Scheibner} These structures allow studies of different mechanisms of coherent optoelectronic control of excitonic states in the pursuit of stable and well characterized qubits.\cite{Control} A fundamental step in this direction is the identification of the relevant interdot quantum couplings. Recently, level anticrossing spectroscopy (LACS) experiments performed on InGaAs/GaAs QDM samples,\cite{Stinaff,Scheibner} have beautifully demonstrated the nature of the dressed spectrum and exciton molecular resonances where charge inter-dot tunneling is the relevant quantum coupling. Tunneling induced anticrossing optical signatures were found as function of applied electric field and pump laser frequency; each anticrossing originating from the mixing between direct and indirect Stark shifted exciton levels.\cite{Szafran} Similarly, investigations on nanocrystal QDs \cite{FRET_Exps_Nano,Amina} and self-assembled QD arrays, \cite{FRET_Exp_Self} have shown that the F\"{o}rster resonant energy transfer mechanism (FRET), \cite{FRET} may be important as well in the full determination and control of interdot coupling, with potential applicability on the development of quantum logic gates.\cite{Nazir} Therefore, a natural question is what kind of FRET signatures could arise in a general LACS experiment, despite the strong effects of charge tunneling on the QDM behavior.

In this work, we investigate the optical signatures of FRET on the dressed spectrum of a tunnel coupled QDM under laser excitation and applied static field. We perform numerical simulations that incorporate realistic input parameters, such as electron and hole tunnelings and interband transition moments, extracted from recent experiments \cite{tunnelings,dipoles,Parameters,mistmatch} or calculations. \cite{Bester} Our work indicates that the strength of FRET optical signatures {\em in the presence of large charge tunneling}, are controllable by laser-exciton detuning, exciton DC stark shift and excitation power. This opens the possibility of observing FRET signatures on the electric field dependent optical spectrum of realistic QDMs. In Sec.\ \ref{sec:Model} the model for the QDM is introduced. An excitonic population map, as function of applied bias and pump laser energy is constructed, resembling the dressed LACS spectra of the QDM. Our main results are presented for two different cases. In Sec.\ \ref{sec:singleresults} we consider the LACS for a QDM dressed spectrum involving F\"{o}rster coupling of direct single excitons.  We show how the coupling is manifested in the variation of  the amplitude (brightness) of a FRET split satellite as a function of the applied electric field: strong at high fields, where the direct-indirect exciton resonance is negligible, decaying as tunneling becomes becomes important, and vanishing completely at the molecular resonance (anticrossing). In Sec.\ \ref{sec:biexcitonresults} we discuss the results of the dressed spectrum including biexcitonic effects, which arise generally upon strong excitation. Although the level manifold including biexcitons is much more complex, we confirm that the signatures do not change qualitatively with respect to those of the simpler model in Sec.\  \ref{sec:singleresults}.  In Sec.\ \ref{lastsec} we consider the F\"{o}rster coupling between spatially indirect biexciton complexes, which results in signatures appearing as split-off satellites on these trion-like biexciton states. The satellite and the splitting are quite resilient against variations of electric field over a wide range, a unique signature which should facilitate experimental identification.

\section{Model}
\label{sec:Model}

The QDM consists of nonidentical ``top" (T) and ``bottom" (B) quantum dots, see Fig.\ 1a, separated by a barrier of width $d$.  The QDM is under an axial electric field $F$ by the application of a top gate voltage. This is usually achieved by placing the QDM in an $n$-$i$ Schottky junction.  In what follows, we denote excitons by $_{h_{B}h_{T}}^{e_{B}e_{T}}X$, where $e_{B(T)},h_{B(T)}=\lbrace 0,1,2\rbrace$ are the electron and hole occupation numbers on the B (T) QD, resulting in a total of 14 neutral exciton states, which correspond to all possible combinations up to double occupancy of electron and hole levels, specifically: the vacuum $_{00}^{00}X = | 0 \rangle$; two single {\em direct} exciton states $_{10}^{10}X$ (bottom exciton) and $ _{01}^{01}X$ (top exciton); two single {\em spatially indirect} exciton states $_{01}^{10}X$ and $_{10}^{01}X$; direct biexciton states $_{20}^{20}X$ and $_{02}^{02}X$; indirect biexciton states $_{02}^{20}X$ and $_{20}^{02}X$; the delocalized biexciton $_{11}^{11}X$, and a set of trion-like indirect states: $_{11}^{02}X$, $_{11}^{20}X$,$_{02}^{11}X$, and $_{20}^{11}X$.  The QDM is continuously pumped by a broad square laser pulse of frequency $\omega$, which in general excites different nearby exciton levels.  The pulse duration is long enough to capture several amplitude oscillations of the excitonic populations.

Complex molecular states arise in this system from the interplay between charge tunnelings, $t_{e}$ and $t_{h}$, the incident radiation field, and intra- and inter-dot Coulomb interactions. The latter are the basic origin of FRET, \cite{FRET} where a (near) resonant interdot dipole-dipole interaction, $V_{F}$,  allows a {\em donor} QD to transfer its exciton energy to the {\em acceptor} neighboring dot, effectively resulting in the
non-radiative interdot ``hoping" of the exciton. Specifically,
\begin{equation}
V_F=\frac{{\mu}_T \, {\mu}_B}{4\pi\epsilon_0\epsilon_{r}d^{3}} \kappa \, ,
\end{equation}
where $\epsilon_{r}$ is the dielectric constant, and $\mu_{T(B)}\sim 6.2 e${\AA} are the interband transition dipole moments, \cite{dipoles} which have been characterized extensively. \cite{dipoles,Parameters,Bester} The FRET strength depends on the size and relative orientation of the dipole moments, which we assume are parallel to each other and perpendicular to their separation, $\kappa\sim 1$.  This gives a value of $V_F=0.08$meV at $d=8.4$nm, consistent with values used in recent theoretical work.\cite{Nazir}  Notice that the FRET rate, $K_{DA}= \frac{2 \pi}{\hbar}V_{F}^{2} \Theta$, depends on the spectral overlap $\Theta$ between the donor emission and acceptor absorption.\cite{Amina,FRET,Lakowicz} Narrow broadenings are characteristic of exciton levels in these systems, which makes FRET relatively rare.  However, the mismatch of the bottom-top donor-acceptor levels, $\Delta_{X_{TB}}$, depends on structural and strain parameters, and may result in a near resonant regime depending on growth conditions.\cite{mistmatch}  For structures such that $\Delta_{X_{TB}}\le V_F$, the coupling is strong and the transfer corresponds to nearly coherent ``exciton hopping;" in contrast, in the weak coupling limit, where $\Delta_{X_{TB}}\gg V_F$, the transfer is typically incoherent, phonon assisted via an auxiliary excited level.\cite{Leegwater}

\begin{figure}[h]
\includegraphics[totalheight=1.0\columnwidth,width=1.0\columnwidth]{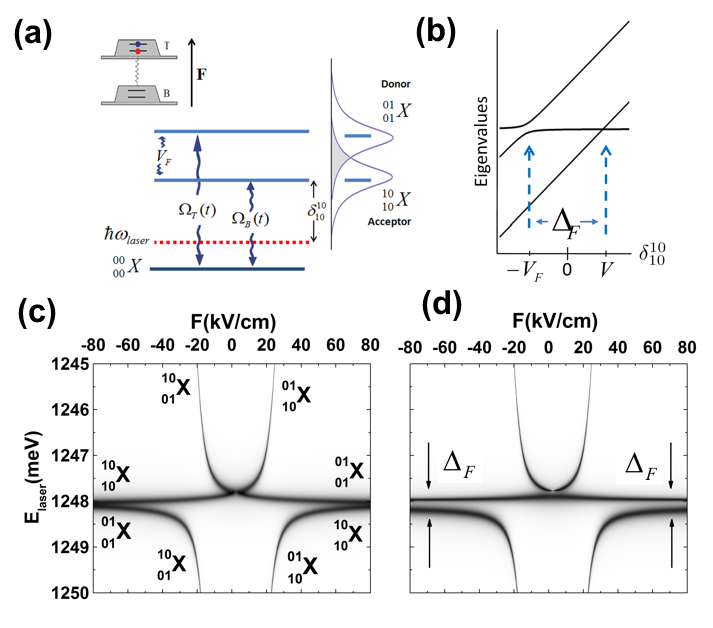}
\caption{(Color online) (a) Single-exciton energy level diagram showing the relevant couplings with the empty QDM state, $|0\rangle$. (b) Dressed spectrum as function of energy detuning  $\delta_{X_{10}^{10}}$. (c) and (d) Show occupation maps of vacuum state $|0\rangle$ in Eq.\ (2), exhibiting features as indicated, where direct,$_{10}^{10}X$, $_{01}^{01}X$, and indirect excitons, $_{10}^{01}X$, $_{01}^{10}X$, have non-zero occupation; for vanishing (in c) and non-zero  $V_{F}$ (d).
Notice sizeable $\Delta_F$ splitting away from tunneling anticrossings in (d).}
\end{figure}

For a given laser frequency $\omega$, and gate field $F$, the exciton dynamics is given by the Hamiltonian, ${H}={H}_{0}+{H}_{\Omega}+{H}_{T}+{H}_{F}$, represented in the excitonic basis listed above $\lbrace\vert i\rangle=\vert_{h_{B}h_{T}}^{e_{B}e_{T}}X\rangle$, $i=1\ldots 14 \rbrace$. In the rotating wave approximation, \cite{RWA}  ${H}_{0}=\sum_{i} \delta_{i}\vert i\rangle\langle i \vert$, contains the detunings of the exciton levels from the laser energy, $\delta_{i}=\epsilon_{i}-\hbar\omega$. ${H}_{\Omega}=\sum_{ij} (\Omega_{ij}\vert i \rangle\langle j \vert + h.c. )$  is the interaction with the radiation field, where $\Omega_{ij}=\langle i \vert \vec{\mu}\cdot\vec{E}\vert j \rangle/\hbar$.
${H}_{T}= \sum_{ij} (t_{ij}\vert i \rangle\langle j \vert + h.c.)$ contains the {\em charge} tunneling matrix elements and ${H}_{{F}}= \sum_{ij} (V_{ij}\vert i \rangle\langle j \vert + h.c.)$ is the F\"orster interaction which describes exciton ``hopping" in the QDM.

The  $\lambda_{j}$ eigenvalues of ${H}$ give the ``dressed'' spectrum of the system,\cite{Cohen} with eigenstates $\vert\Psi_{l}\rangle=\sum_{i}b_{i}^{l}\vert i\rangle$, which are mixtures of the bare exciton states $| i \rangle$. The effective Rabi period, $T_{R}$,  is function of all exciton couplings of the Hamiltonian and is of the order of a few ps, \cite{Rabi} so that $T_R \ll \tau_X \sim 1ns$,  the radiative exciton recombination time.\cite{lifetimes} In the strong coupling limit,\cite{Leegwater} we assume $V_F\gg\Delta_{X_{TB}}, \hbar/\tau_X$, so that
the dynamics of the system is approximately coherent for times $t$, such that, $T_R < t \ll \tau_{X}$. In the following, for simplicity, we report the results of a fully coherent (Hamiltonian) evolution, although introduction of different relaxation and incoherent processes, \cite{VillasNoMarkov,NakaokaFinley} via a Lindblad formulation of the density matrix evolution, results in qualitatively similar results, \cite{Enrique-Italy} and does not affect our main conclusions. In this regime, the time evolution is unitary, given by $U(t)=\exp(\frac{-i{H}t}{\hbar})$, so that the population of an exciton state $\vert i\rangle$ is given at time $t$ by $P_{i}(t)=\vert\langle i\vert U(t)\vert 0\rangle\vert^{2}$, starting from an `empty' QDM in state $|0\rangle$.  The average population for long integration times is given by $p_i = (1/t_{\infty})\int_{0}^{t_{\infty}}P_{i}(t)dt$, where $t_\infty$ stands for a constant amplitude pulse duration long enough to capture several amplitude oscillations of the exciton populations, $T_R \ll t_\infty < \tau_X$.  We find that only a few Rabi oscillations are necessary to compute the long-time average population after transient effects, and well before a time scale where amplitude damping may become relevant.

\section{Results: single excitons}
\label{sec:singleresults}

Let us consider the effect of F\"orster coupling of near resonant top $_{01}^{01}X$ and bottom $_{10}^{10}X$ direct exciton levels, shown on the qualitative level diagram in Fig.\ 1a. For QDs coupled by tunneling, it is important to include the single indirect exciton states.\cite{numeros}
Although one can study the dynamics of the system using the entire basis of neutral excitons (the 14 states described above and used in Fig.\ 2b,d,f {\em et seq}.\ below), it is simpler and more intuitive to consider only the Hamiltonian representing only the single-exciton states in the basis, both with direct and indirect character,
\begin{equation}
{H} = \left(
\begin{array}{ccccc}
\delta_{0} & \Omega_{T} & 0 & 0 & \Omega_{B} \\
\Omega_{T}& \delta_{01}^{01} & t_{e} & t_{h} & V_{F} \\
0 & t_{e} & \delta_{01}^{10}+\Delta_{S} & 0 & t_{h} \\
0 & t_{h} & 0 & \delta_{10}^{01}-\Delta_{S} & t_{e}\\
\Omega_{B} & V_{F} & t_{h} & t_{e} & \delta_{10}^{10}
\end{array}
\right) \, , \label{eq2}
\end{equation}
where the columns are associated with the states $|0\rangle$,
$|_{01}^{01}X \rangle$, $|_{01}^{10}X \rangle$, $|_{10}^{01}X \rangle$, and
$|_{10}^{10}X \rangle$, so that the third and fourth columns represent spatially indirect excitons, and $\Delta_s=eFd$ is their Stark shift.

We can simulate a generic LACS map from the average population of different exciton states, as function of the laser pump energy and applied axial electric field.   These average population maps for a given coordinate $(F,\hbar\omega)$ are obtained from the dynamics of the QDM and projected back to the original exciton basis. Then, any exciton state populated under pumping will exhibit a relative amplitude $p_i$ and a feature on the corresponding map.  In general, two or more excitons will share population at a given $(F,\hbar\omega)$ coordinate if they have non-vanishing components in a dressed state; then by examination of maps corresponding to individual excitons, one can reconstruct the entire dressed spectrum of the system. Alternatively, one can compute the population map of the vacuum state $_{00}^{00}X=|0\rangle$, so that the complete dressed LACS spectrum will correspond to all $(F,\hbar\omega)$ coordinates where this estate is {\em depopulated} (and therefore there are non-vanishing amplitudes at the various exciton states of the QDM).

A LACS population map for $|0\rangle$ corresponding to the simplified Hamiltonian in Eq.\ (2) is shown in Fig.\ 1c, for $d=8.4$ nm, assuming that $_{10}^{10}X$ and $_{01}^{01}X$ have negligible F\"{o}rster coupling, so that FRET is turned off ($V_F=0$). In this case, the only interdot couplings are the electron and hole tunnelings, $t_{e}$ and $t_{h}$; notice also that the only mixing between direct exciton lines is due to radiation, $\Omega_{T(B)}$.
Through the Stark effect, the applied field $F$ shifts the indirect excitons by $\pm\Delta_{S}$, tuning them into resonance with the direct excitons.  Near resonance ($F\simeq \pm 20$kV/cm in this case), the charge tunneling produces mixing between the direct and indirect excitons, with the resulting anticrossings clearly visible in the exciton population maps.  For simplicity we have neglected the confined Stark effect which would also shift the direct excitons slightly, as this is not an essential element in our discussion.

However, if the direct excitons are near resonant and strongly coupled by $V_F$, FRET would have the effect of considerably splitting the direct exciton lines by $\Delta_F$, Fig.\ 1d (for $\hbar \omega \simeq 1248$ meV, although mixings near the tunneling anticrossing are dominated by $t_e$).
The F\"orster coupling signature is seen more clearly in Fig.\ 2a, which shows the average population of the acceptor exciton $_{10}^{10}X$. The splitting of the exciton line near $\omega \simeq 1248$ meV appears clearly as a non-dispersing satellite line away from the anticrossing. Remarkably, for $F \simeq (-35, -20)$ kV/cm, the satellite disperses away, following the asymptotic indirect line of $_{01}^{10}X$. It is interesting that the FRET mixing (in conjunction with electron tunneling), makes this indirect exciton acquire some of the amplitude and ``light up." The FRET signature eventually bleaches at the anticrossing where tunneling dominates (at low electric fields), explicitly demonstrating that tunneling is detrimental to FRET.\@  We emphasize that a proper description of the dynamics of such system needs to take into account the entire set of exciton states, since the direct coupling terms in the Hamiltonian and the various higher order virtual processes make the decoupling of direct and indirect exciton subspaces not possible.

\begin{figure}[t]
\includegraphics[totalheight=1.0\columnwidth,width=1.0\columnwidth]{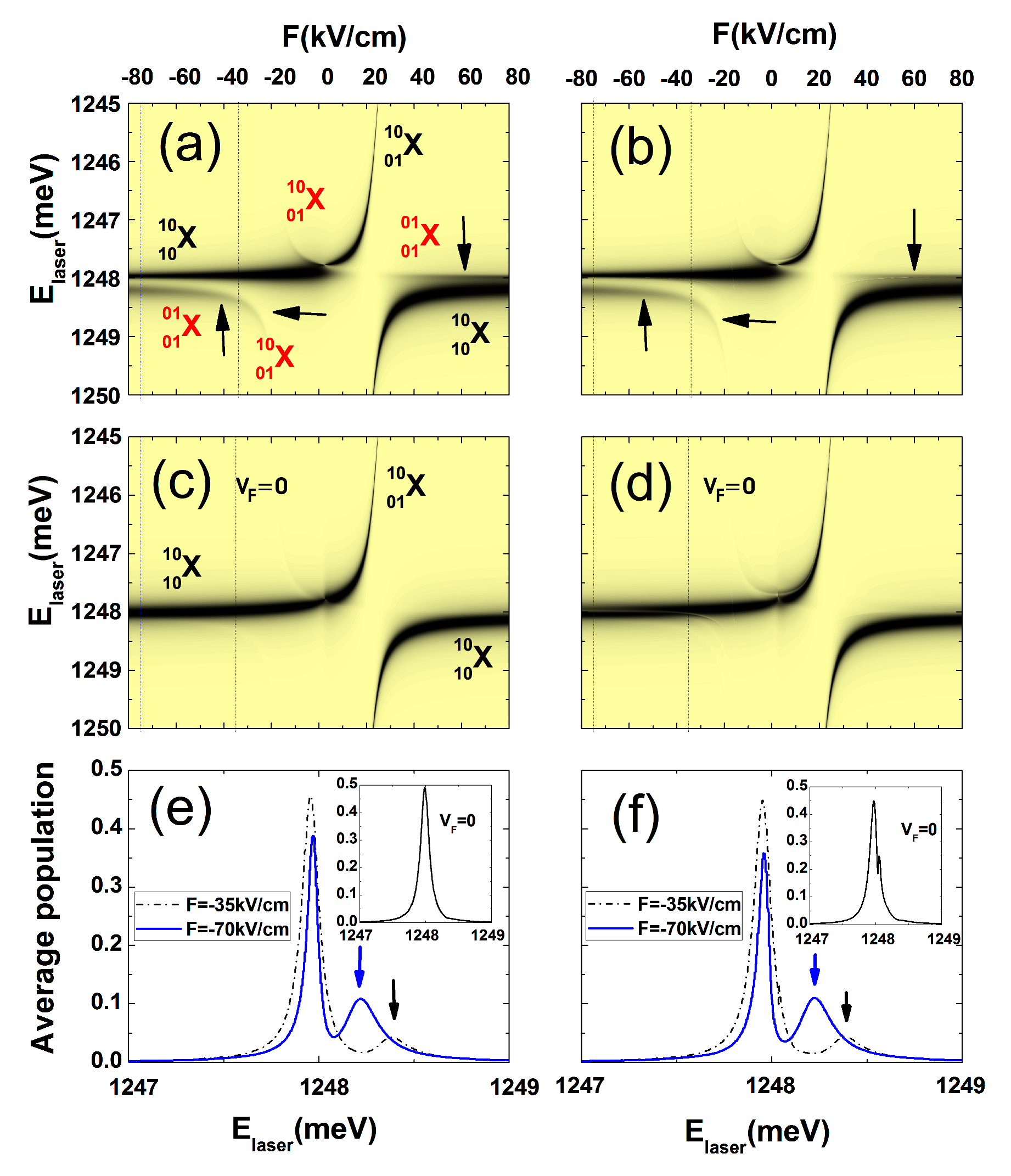}
\caption{(Color online) (a) Population map of {\em acceptor} exciton $_{10}^{10}X$ for system described by Eq.\ (\ref{eq2}) shows FRET-induced weak satellite line (black arrows), parallel up to tunneling anticrossing region. (b) Signature persists, even when taking all biexciton transitions into account (all 14 states). The respective population maps (c) and (d) show no satellites for $V_F=0$ (no FRET).  (e) Acceptor population for two fixed values of $F=-35$ (black dashed line) and $F=-70$kV/cm (blue solid). (f) Acceptor population as in (e) but for full system in (b). Notice absence of features when $V_F=0$ in insets.}
\end{figure}

Figure 2e compares the populations for fixed electric field values, $F=-70$ and
$-35$kV/cm. The FRET satellite peak is higher at stronger fields; to understand this behavior, we reduce the Hamiltonian onto a subspace that mixes only the two direct excitons,
$\vert _{10}^{10}X \rangle,\vert _{01}^{01}X\rangle$, with the empty QDM, $|0\rangle$,
\begin{equation}
{H}_{D} = \left(
\begin{array}{ccc}
0 & \Omega_{T} & \Omega_{B} \\
\Omega_{T}& \delta_{01}^{01} & V_{F} \\
\Omega_{B} & V_{F} & \delta_{10}^{10}
\end{array}
\right) \, .
\end{equation}
For $\Omega_{B}\simeq\Omega_{T}$ and $\delta_{10}^{10}\simeq\delta_{01}^{01}$, the dressed spectrum of this Hamiltonian, shown as function of $\delta_{10}^{10}$ in Fig.\ 1b,  reveals two anticrossings, one with with large gap at $\delta_{10}^{10}=-V_{F}$, and another with negligible gap at $\delta_{10}^{10}=(V_{F}^{2}-\Omega_{B}^{2})/V_{F} = V$, and with a relative separation $\Delta_F=2V_F-(\frac{\Omega^{2}}{V_F})$.
Therefore, when the laser energy is set to $\hbar\omega\approx E_{_{10}^{10}X}+V_{F}$, the system is strongly coupled by $V_{F}$, and a strong population of the acceptor exciton $_{10}^{10}X$ is noticeable; however, when $\hbar\omega\approx E_{_{10}^{10}X}-V$, the coupling is weak and the population low.  Here V can be positive or negative depending on the relative magnitude of $V_F$ and $\Omega$. Therefore, the acceptor exciton line would split into a strong line separated by $\Delta_F$  (for $|F| \gg 1$) from a weak and narrow satellite. For low excitation power, $\Omega^2\ll V_F$, a better F\"{o}rster splitting resolution is achievable as $\Delta_F\simeq 2V_F$.  Notice splitting in Fig.\ 1d is $\simeq 0.25$ meV ($\gtrsim \Delta_F=0.14$meV), as the gap is magnified slightly by power broadening and electron tunneling corrections even for $F \simeq -80$kV/cm. This effect arises from tunneling correction terms to the non-diagonal matrix elements of Eq.\ (3) at intermediate electric fields, for which Eq. \ (2) or the full Hamiltonian is more appropriate.
If the donor/acceptor excitons are in resonance, $\delta_{01}^{01} \simeq \delta_{10}^{10} $, the time-dependent exciton populations are given by
\begin{equation}
\vert\psi_{10}^{10}(t)\vert^{2}=\vert\psi_{01}^{01}(t)\vert^{2}= \left(\frac{2\Omega^{2}}{\Omega_{R}^{2}}\right) \left(1-\cos(\Omega_{R}t) \right) \, ,
\end{equation}
where $\Omega_{R}=\sqrt{(\delta_{10}^{10}+V_{F})^{2}+8\Omega^{2}}$ is the Rabi frequency at that detuning [notice that the eigenvalue gap at $\delta_{10}^{10} = - V_{F}$ in Fig.\ 1b is $2\sqrt{2} \Omega$], and $\Omega=\Omega_T=\Omega_B$. Correspondingly, the  average exciton population is given by $p_{10}^{10}=2\Omega^{2}/\Omega_R^{2}$, which is clearly maximal for $\delta_{10}^{10}=-V_F$, and monotonically increases with laser power for other detunings. The tunability of $p_{10}^{10}$ with laser detuning and excitation power, suggests control of the FRET effective coupling via laser parameters, for a given QDM, which may be important in the experimental identification and utilization of this effect.

\begin{figure}[htb]
\includegraphics[totalheight=1.0\columnwidth,width=1.0\columnwidth]{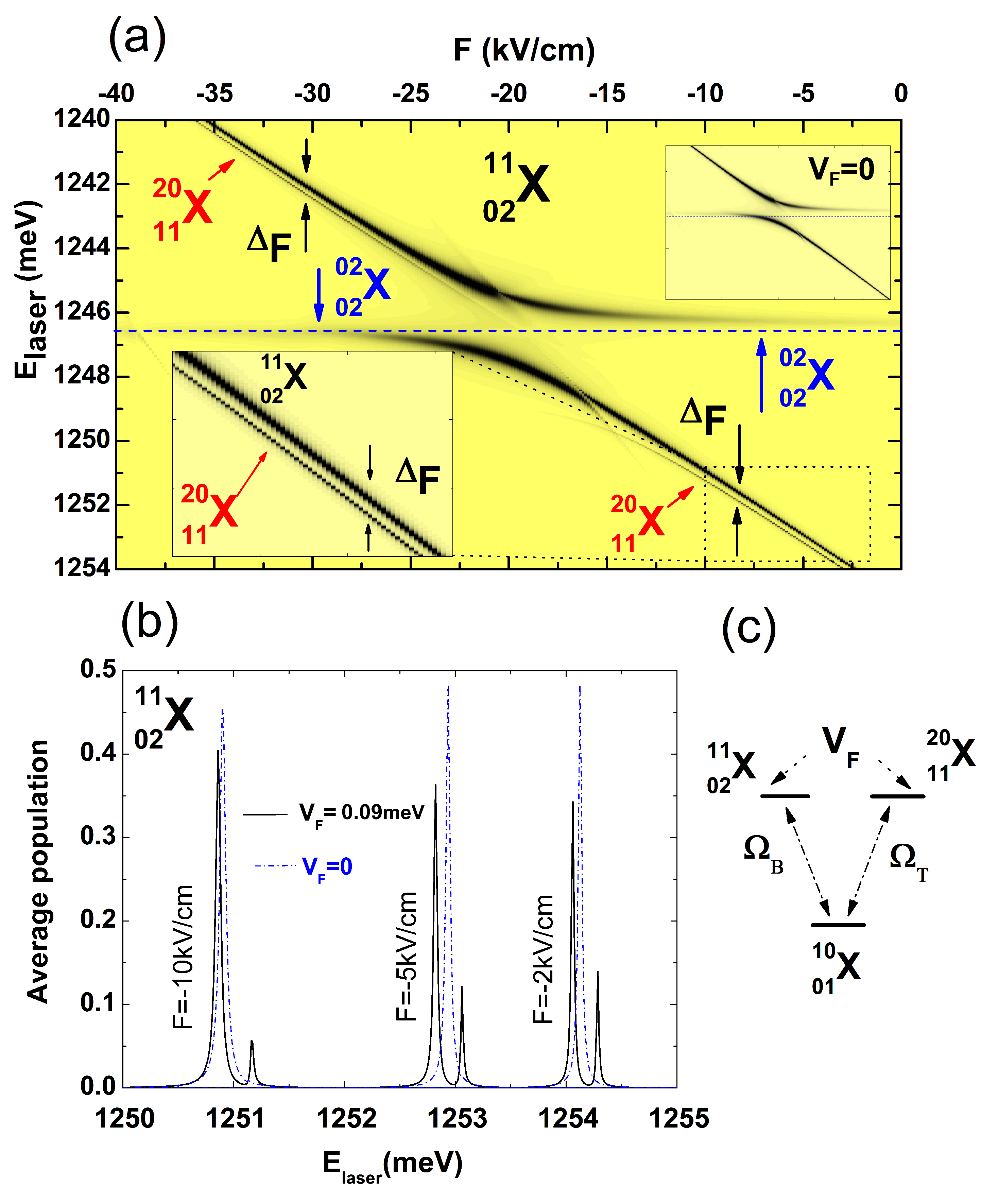}
\caption{(Color online) (a) Population map of {\em acceptor} exciton $_{02}^{11}X$, showing splitting of indirect exciton line over large $F$ range. Blue dashed line indicates the asymptotic position of the direct biexciton $_{02}^{02}X$ mixing with the acceptor level. Top right inset: map with no FRET. Bottom left inset: Zoom of indirect line splitting. (b) Average population for three values of $F$, showing invariance of indirect split-off energy: satellite line in (a) tracks the brighter exciton line nearly parallel. (c) Reduced level diagram describing the system for high values of $F$.}
\end{figure}

\begin{figure}[htb]
\includegraphics[totalheight=1.0\columnwidth,width=1.0\columnwidth]{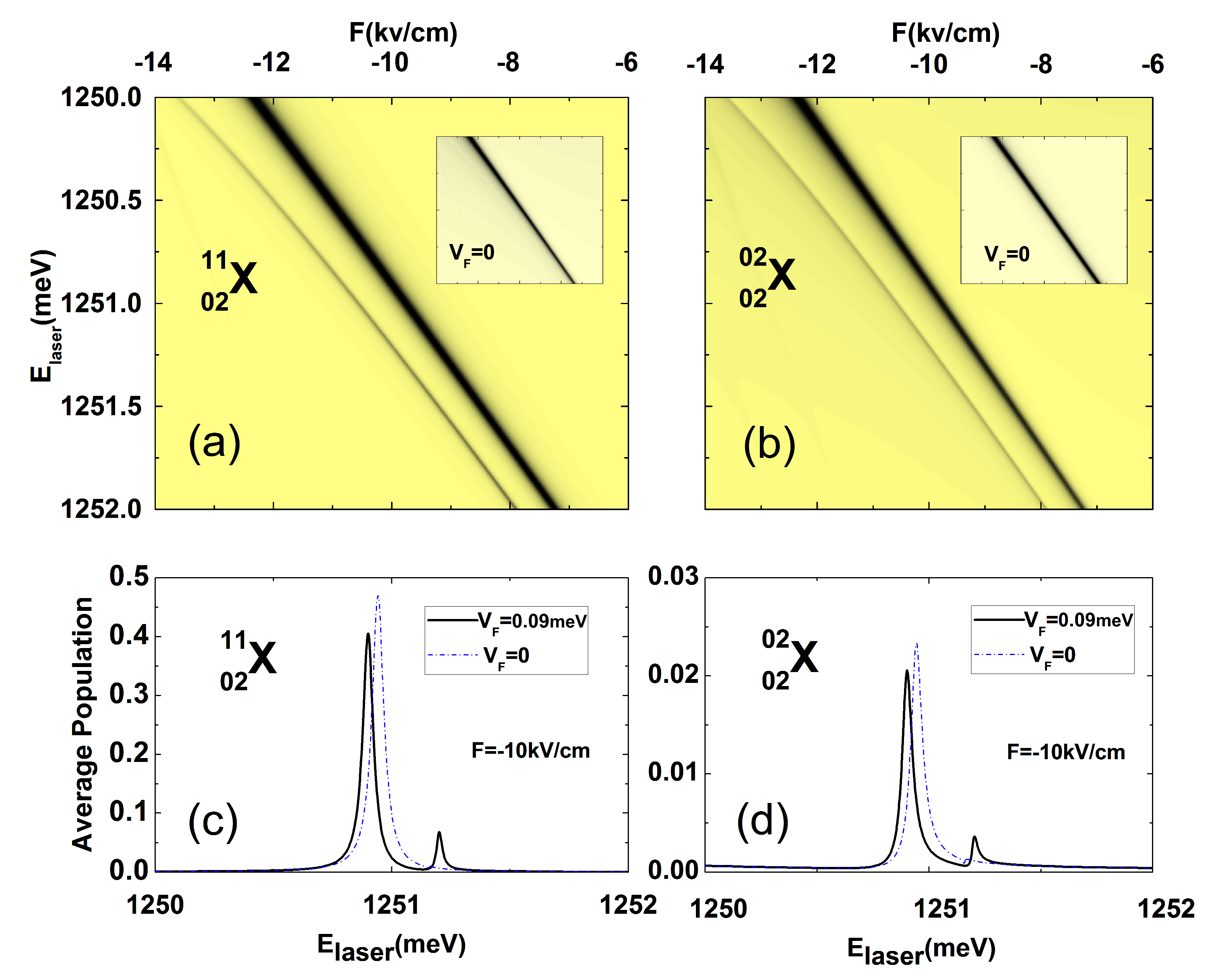}
\caption{(Color online) Splitting under FRET. (a) The trion-like state $_{02}^{11}X$ has satellite after mixing with the biexciton resonance $_{02}^{02}X$, as shown in (b). Notice inset shows no splitting when $V_F=0$. Comparison of (c) and (d) confirms mixing of FRET satellites between direct and trion-like biexciton states; FRET signature on direct biexciton state is weaker by an order of magnitude.}
\end{figure}

\subsection{Biexcitonic effects}
\label{sec:biexcitonresults}

Under strong laser excitation,\cite{Strong} there is the possibility of exciting additional exciton levels outside the relevant subspace of consideration.
For example, excited levels of a direct (indirect) exciton appear usually at a few meV higher energy, as they correspond to excited levels of the electron or the hole, depending on the value of the effective mass.\cite{Stinaff,Scheibner,Parameters}
Other excitations, such as LO phonon resonances, appear $\sim 35$meV above the lowest exciton transition for GaAs,\cite{Semiconductor} and can be safely ignored.
However, biexciton resonances cannot be usually ignored in the dynamics of single excitons of Eq.\ (2), as their detuning is at most a few meV.\cite{Biexciton} The pumping of biexcitons in either QD, $_{20}^{20}X$, $_{02}^{02}X$, may then be expected to strongly affect the dynamical evolution of the single exciton states, as the additional subspace spanned by all biexcitonic complexes would extract population from the single exciton states. However, the FRET split-off satellite is remarkably persistent and appears with significant intensity.  In Fig.\ 2f we present the results of the full Hamiltonian (14 states), showing only a slight attenuation of the main peak, while the satellite remains strong.  The differences with Fig.\ 2e arise from mixing with an increased number of exciton states, and interestingly results in better contrast of the F\"orster signature.

\section{FRET in biexcitonic complexes}
\label{lastsec}

Let us consider a different FRET process involving partial spatially indirect biexciton complexes, $_{02}^{11}X$, $_{11}^{20}X$.  These states are naturally accompanied by other nearby indirect biexcitons,  $_{11}^{02}X$ and $_{20}^{11}X$,  and as all carry a large effective dipole moment, they would be efficiently coupled by FRET if in/near resonance.  These {\em neutral} states are trion-like objects, as they can be pictured as a trion in one dot, \cite{trions} plus a residual charge in the second dot.
Notice that $_{02}^{11}X$, $_{11}^{20}X$ experience the same Stark shift $\Delta_S = eFd$, while $_{11}^{02}X$, $_{20}^{11}X$ shift with opposite slope by $-\Delta_S$.  Therefore, for large $\Delta_S$ and $\delta_{02}^{11}$ similar to $\delta_{11}^{20}$, the evolution of such system can be studied by considering the effective three-state Hamiltonian
\begin{equation}
{H}_{I} = \left(
\begin{array}{ccc}
\Delta_{S} & \Omega_{T} & \Omega_{B} \\
\Omega_{T}& \delta_{02}^{11} + \Delta_{S} & V_{F} \\
\Omega_{B} & V_{F} & \delta_{11}^{20} + \Delta_{S}
\end{array}
\right) \, ,
\end{equation}
where FRET couples $_{11}^{20}X$ and $_{02}^{11}X$, while the indirect exciton $_{01}^{10}X$ (first column) couples directly to these biexciton states via the laser field, as Fig.\ 3c indicates. Since all three levels acquire the same Stark shift $\Delta_{S}$,  the eigenvalues are invariant against $F$ variation. This symmetry makes Eq.\ (5) similar to (3), so that the FRET coupling should result in a bright exciton line with a split-off satellite that follows an indirect exciton line. Our simulations for the full Hamiltonian with 14 states confirm that this occurs for a wide range of $F$, see Fig.\ 3a, except at the tunneling anticrossings where the trion-like exciton mixes strongly with the direct exciton $_{02}^{02}X$. Moreover, the splitting itself does not change appreciably with $F$, see Fig.\ 3b, even as the amplitude varies.
The population intensity increases as one moves away from the anticrossing, since the exciton population is shared only among three states.
The strong mixing with $_{02}^{02}X$ in the range of field $F$ shown results in `dressing' from the remaining direct exciton lines at $\hbar\omega\sim1251$meV, makes the biexcitonic FRET  signature appear as a weak satellite split peak, see Fig.\ 4b. Figures 4c and 4d compare the much different population amplitudes of the trion-like and direct biexcitons for the same region of laser frequency.

\section{Concluding remarks}
\label{sec:Conclusion}

In summary, we have shown it is possible to detect FRET signatures on the level anticrossing dressed spectrum of QDMs, with strength controllable by laser detuning and excitation power, despite the dominant effect of electron (hole) tunneling. The direct acceptor exciton line splits, so that away from the tunneling anticrossing (large Stark shift), a satellite peak follows the acceptor direct line. Near the anticrossing, however, the line follows an indirect exciton line that mixes with the direct exciton donor level, vanishing completely at the anticrossing, where tunneling dominates. FRET among trion-like indirect biexcitons generates signatures that also appear as a split exciton line and are robust over a wide range of Stark shifts.

The use of differential transmission spectroscopy measurements of exciton populations is well suited to investigate the dressed spectrum of QDs.\cite{Jundt} An excitonic state is populated in the first dot via pulsed pump laser excitation. A second weak probe pulse could be sent into resonance with the excitonic transition frequency of the second dot (which is different from the first dot), then measure the transient differential transmission of this probe, reflecting its population. For example, for single exciton direct transitions, if an electron is in the second dot, the probe photon cannot be absorbed there because of Pauli blockade (if both pump and probe carry the same polarization). Moreover, in this technique lineshape broadening may be suppressed further by the careful interplay of the laser excitation and probe pulses.

As these FRET signatures become identified in experiments, they may be useful as an interesting alternative probe and control of interdot interactions and the quantum state of exciton-defined qubits.  We look forward to further experimental and theoretical studies in this direction.

\acknowledgments

We thank J. M. Villas-Boas and E. Stinaff for helpful discussions, and support from NSF-DMR MWN grant 0710581, and OU BNNT and CMSS.


\begin{thebibliography}{99}
\bibitem{numeros} Exciton transition energies and structure parameters, consistent with [\onlinecite{Bester}], and  derived from measurements [\onlinecite{tunnelings}] are as follows. Figures 1c,1d,2: interdot distance $d\simeq8.4nm$, $t_{e}, t_h= 2.0, 0.1$meV, $V_F=0.08$meV, $\Omega_{T(B)}=40$$\mu$eV. Exciton bare energies: $E_{_{01}^{01}X}=1248.01$meV, $E_{_{10}^{10}X}=E_{_{01}^{01}X}+10\mu$eV. Figures 3,4: $d\sim8.25nm, t_{e},t_h=2.1,0.12$meV, $V_F=0.09$meV, $\Omega_{T(B)}=140\mu$eV. Biexciton bare energies: $E_{_{02}^{02}X}=2492.85$meV,$,E_{_{11}^{02}X}=E_{_{02}^{02}X}+17$meV, and $E_{_{20}^{11}X}=E_{_{11}^{02}X}-20\mu$eV. Interband transition moments $\mu_{T(B)}=6.2e${\AA} [\onlinecite{dipoles}]. \@ Exciton-radiation couplings (controlled by the laser pulse amplitude) are in the strong pumping regime, $\Omega_{T(B)}\gg\tau_{X,(XX)}$. Exciton (biexciton) radiative recombination times $\tau_{X} (\tau_{XX})\sim 1.0\, (0.5)$ns [\onlinecite{lifetimes}].
\end{thebibliography}
\end{document}